\documentclass[reprint,amsmath,amssymb,pra,aps,twocolumn,showpacs]{revtex4-1}

\usepackage{graphicx}
\usepackage{subfigure}
\usepackage{dcolumn}
\usepackage{bm}
\usepackage{color}

\def\hatd#1{\hat{#1}^\dagger}
\def\bra#1{\left\langle{#1}\right|}
\def\ket#1{\left|{#1}\right\rangle}
\def\braket#1#2{\left\langle{{#1}}\mathrel{\left|{\vphantom{{#1}{#2}}}\right.\kern-\nulldelimiterspace}{{#2}}\right\rangle}

\def\RomanNum#1{\textrm{\uppercase\expandafter{\romannumeral#1}}}

\begin{document}

\title{Heisenberg-limited Sagnac Interferometer with Multi-particle States}

\author{Chengyi Luo$^{1,2}$}

\author{Jiahao Huang$^{1}$}
\altaffiliation{Email: hjiahao@mail2.sysu.edu.cn, eqjiahao@gmail.com}

\author{Xiangdong Zhang$^{1,2}$}

\author{Chaohong Lee$^{1,3}$}
\altaffiliation{Email: lichaoh2@mail.sysu.edu.cn, chleecn@gmail.com}

\affiliation{$^{1}$TianQin Research Center \& School of Physics and Astronomy, Sun Yat-Sen University (Zhuhai Campus), Zhuhai 519082, China}

\affiliation{$^{2}$School of Physics, Sun Yat-Sen University (Guangzhou Campus), Guangzhou 510275, China}

\affiliation{$^{3}$State Key Laboratory of Optoelectronic Materials and Technologies, Sun Yat-Sen University (Guangzhou Campus), Guangzhou 510275, China}

\date{\today}

\begin{abstract}
  The Sagnac interferometry has been widely used to measure rotation frequency.
  Beyond the conventional single-particle Sagnac interferometry, we propose an atomic Sagnac interferometry via multi-particle entangled states.
  In our scheme, an ensemble of entangled two-state Bose atoms are moved in a ring by a state-dependent rotating potential and then are recombined for interference via Ramsey pulses after a specific time determined by the state-dependent rotating potential.
  The ultimate rotation sensitivity can be improved to the Heisenberg limit if the initial internal degrees of freedom are entangled.
  By implementing parity measurement, the ultimate measurement precision can be saturated and the achieved measurement precisions approach to the Heisenberg limit.
  Our results provide a promising way to exploit many-body quantum entanglement in precision metrology of rotation sensing.
\end{abstract}

\maketitle


\section{\label{sec:Sec1}Introduction}

Various advantages of quantum metrology~\cite{Gross2010,Riedel2010,Lucke2011,Ockeloen2013,Mussel2014,Strobel2014,Vasilakis2015} have been demonstrated by neutral atoms~\cite{Zimmermann2007,Bloch2008,Chin2010}, trapped ions~\cite{Leibfried2003} and photons~\cite{Raimond2001} etc.
Generally speaking, the measurement precision $\Delta\chi$ via $N$ independent particles is imposed by the standard quantum limit (SQL): $\Delta\chi \propto 1/\sqrt{N}$~\cite{Maccone2004}.
However, by utilizing multi-particle entanglement and squeezing, the SQL can be surpassed~\cite{Wineland1992,Kitagawa1993,Maccone2004,Maccone2006,Maccone2011,Lee2012,Huang2014}.
The Greenberger-Horne-Zeilinger (GHZ) state~\cite{Leibfried2004} and the N00N state~\cite{Kok2002} can improve the minimum uncertainty to the so-called Heisenberg limit~\cite{Maccone2004,Maccone2006,Maccone2011}: $\Delta\chi \propto 1/N$.
It has also been demonstrated that, the achievable precision can beat the SQL or even approach the Heisenberg limit by using spin squeezed states~\cite{Gross2010,Riedel2010,Ockeloen2013,Mussel2014}, twin Fock states~\cite{Lucke2011} and spin cat states~\cite{Huang2015}.
Up to now, quantum metrology has been extensively used in high-precision sensing of rotations~\cite{Haine2016,Nolan2016,Taylor2016}, accelerations~\cite{Peters1999}, magnetic fields~\cite{Ockeloen2013,Mussel2014} and gravitational fields~\cite{LIGO1,LIGO2} etc.

Rotation sensing is essential in both fundamental sciences and practical technologies, from determining the Earth's rotation frequency to building gyroscopes for navigation~\cite{Gustavson1997}.
Sagnac effect describes the phase shift accumulation between two counter-propagating waves around a closed path in a rotating frame~\cite{Post1967}.
Based upon the Sagnac effect, Sagnac interferometers for measuring rotation frequency have been realized via ring lasers~\cite{Schreiber2010}, atoms~\cite{Alan1997,Gustavson2000,Gauguet2009,Tackmann2012} and trapped ions~\cite{Wes2016}.
Recently, a Sagnac interferometry with a single-atom clock~\cite{Fernholz2015} have been proposed via combing the techniques of state-dependent manipulations and Ramsey pulses.

Beyond single-particle quantum states, it is interesting to investigate how to exploit many-body quantum entanglement in precision measurement of rotation frequency.
On one hand, due to their robust quantum coherence and high controllability, several entangled states of ultracold atoms (in particular Bose condensed atoms) have been generated in experiments~\cite{Gross2010,Riedel2010,Ockeloen2013,Mussel2014}.
On the other hand, ring traps and state-dependent manipulation of Bose condensed atoms~\cite{Treutlein2004,Schumm2005,Morizot2006,Fernholz2007,Sherlock2011,Szmuk2015} have been demonstrated.
Combing these techniques, it is possible to realize the Sagnac interferometer with multi-particle entangled states of Bose condensed atoms.
Unlike other conventional interferometers, the external and internal degrees of freedom couple with each other during the phase accumulation in our Sagnac interferometry.
Thus, the measurement precision of rotation frequency is sensitively affected by both the estimated angular frequency itself and the induced angular frequency.
To achieve the best sensitivity, it is important to optimally control the angular frequency.

This article is organized as follows.
In Sec.~\ref{Sec2}, we present our multi-particle Sagnac interferometry scheme.
In Sec.~\ref{Sec3}, we calculate the ultimate rotation measurement precision and find the measurement precision may reach the Heisenberg limit. The uncertainty of the estimated angular frequency $\omega_s$ depends on the induced angular frequency $\omega_p$ as well as the rotation frequency $\omega_s$ to be measured. We also derive an analytic quantum Cramer-Rao bound (QCRB) for some specific choices of $\omega_p$ and $\omega_s$.
In Sec.~\ref{Sec4}, we further investigate the rotation frequency estimation via parity measurement. We find that parity measurement is an optimal and realizable way to obtain the Heisenberg-limited precision. Meanwhile, we also derive an analytic formula for the uncertainty of rotation frequency via parity measurement.
In Sec.~\ref{Sec5}, we briefly summarize and outlook our scheme.

\begin{figure*}[!htb]
\includegraphics[width=180mm]{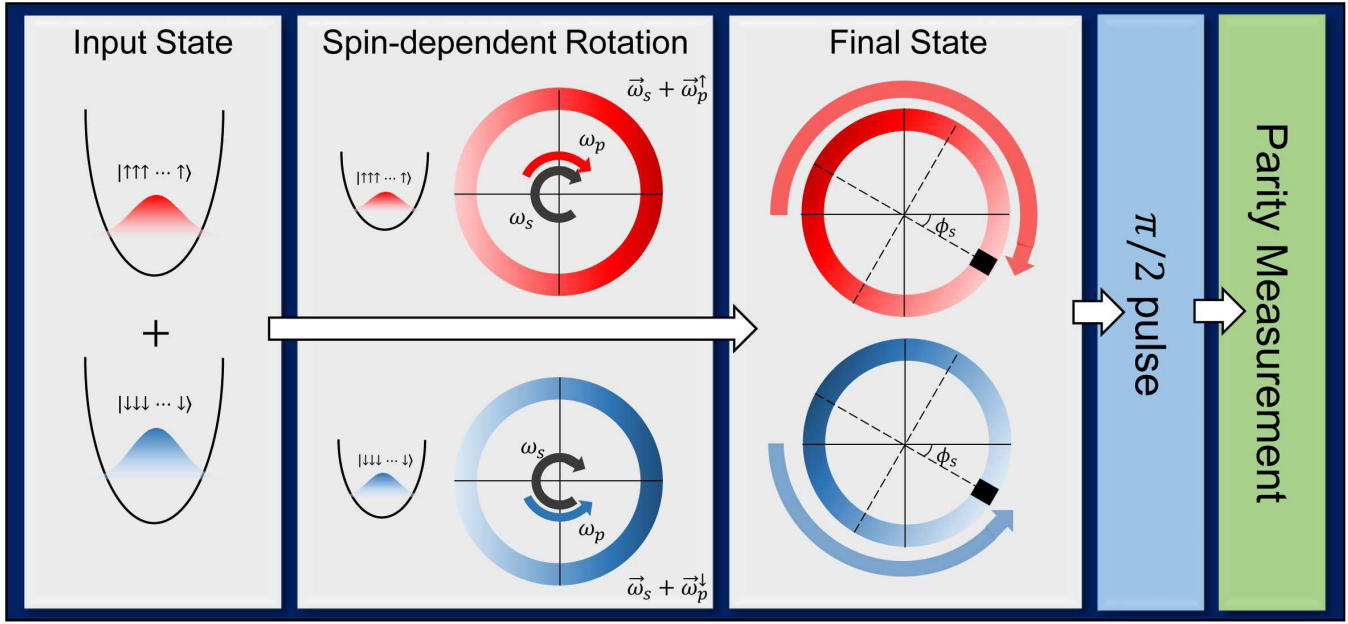}
\caption{\label{Fig_schematic}(color online). Schematic diagram of the multi-atom Sagnac interferometry. Initially, the system is prepared in the internal GHZ state between two hyperfine levels ($|m_F=1\rangle$ and $|m_F=-1\rangle$). Then, a state-dependent evolution is applied for accumulating a relative phase dependent on the rotation angular frequency $\omega_s$. Finally, after a $\pi/2$ pulse for the two hyperfine states, the parity measurement is used to extract the relative phase and the rotation angular frequency $\omega_s$ is derived from the relative phase.}
\end{figure*}

\section{\label{Sec2} Multi-particle Sagnac Interferometry via state-dependent manipulation}

In our scheme, by using Bose condensed atoms, the multi-particle Sagnac interferometry combines state-dependent potentials moving around a ring with a sequence of Ramsey pulses.
The interferometry involves two atomic hyperfine states ($\ket{m_F=1}\equiv\ket{\uparrow}$ and $\ket{m_F=-1}\equiv\ket{\downarrow}$), which label the two evolution paths.
Each atom may occupy one of the two hyperfine states and can be regarded as a spin-$\frac{1}{2}$ particle with $\hat\sigma_z\ket{\uparrow}=+\ket{\uparrow}$ and $\hat\sigma_z|\ket{\downarrow}=-\ket{\downarrow}$.
We assume all the atoms are tightly confined in a ring trap potential of fixed radius $r$ and the motional degrees of freedom is restricted to the azimuthal angle $\theta$.
Initially, all atoms locate at $\theta=0$ and their external states are prepared in the ground state $\ket 0$ of the harmonic potential along the radial direction.

Our multi-particle Sagnac interferometry includes the following steps.
First, a desired multi-particle state is prepared as the input state. For our input state, only the spin degrees of freedom are entangled, while the external degrees of freedom is identically in $\ket 0$.
Then, the spin-dependent trapping potentials for $\ket{\uparrow}$ and $\ket{\downarrow}$ rotate along opposite directions with angular frequency $+\omega_p(t)$ and $-\omega_p(t)$, respectively.
During the free evolution, an $\omega_s$-dependent phase shift $\phi(\omega_s)$ between two counter-propagating modes is accumulated.
Finally, the two modes encounter on the other side and a $\pi/2$ pulse is applied for recombination.
The unknown angular frequency $\omega_s$ is extracted by measuring the population information of the spin states.
The schematic diagram of our multi-particle Sagnac interferometry is shown in Fig.~\ref{Fig_schematic}.

We consider the maximally entangled state (GHZ state) as the input state.
Compared with the input spin coherent state (a multi-particle state without entanglement), the GHZ state can enhance the rotation sensitivity from the SQL to Heisenberg limit.
The input GHZ state is written as
\begin{equation}\label{GHZ}
  \ket{\Psi}_{in}=\frac{1}{\sqrt{2}}\left[\bigotimes^N_{k=1} (\ket{\uparrow}  \ket{0})_k+\bigotimes^N_{k=1}(\ket{\downarrow} \ket{0})_k \right],
\end{equation}
where the internal states of each particle are maximally entangled.
Then the system will undergo the dynamical evolution to accumulate a phase shift between the two spin components.
At this stage, the atom-atom interaction of the Bose condensed atoms can be tuned to zero by some of the techniques, such as Feshbach resonance~\cite{Zimmermann2007}.
The two different spin components rotate in opposite direction with angular velocity $\omega_p(t)$ around the ring trapping potential, which can be described by a spin-dependent Hamiltonian~\cite{Fernholz2015}:
%
\begin{equation}\label{H1}
  \hat{H}(t)=\sum^N_{k=1}\hat{H}_k(t) , \quad \hat{H}_k(t)=\hat{H}_\uparrow(t) \ket{\uparrow} \bra{\uparrow}+ \hat{H}_\downarrow(t) \ket{\downarrow} \bra{\downarrow},
\end{equation}
where $\hat{H}_k$ is the single-particle Hamiltonian for the $k$-th particle containing two parts, $\hat{H}_{\uparrow}(t)$ and $\hat{H}_{\downarrow}(t)$.
In the inertial frame, the explicit expressions for $\hat{H}_{\uparrow}(t)$ and $\hat{H}_{\downarrow}(t)$ are
\begin{eqnarray}\label{HuHd}
  \hat{H}_{\sigma}(t)&=&\hbar \omega \hatd{a}\hat{a}+i\sqrt{\frac{m\hbar \omega}{2}}r\left(\hatd{a}\!-\!\hat{a}\right)\!\left[\omega_s+\eta_{\sigma} \omega_p(t)\right] \nonumber\\
  &=& \hat{H}_0 + \hat{H}_{\sigma}^{I}(t),
\end{eqnarray}
where $\sigma=\uparrow,\downarrow$ and $\omega$ is the trapping frequency of the harmonic potential along the radial direction. $\hatd{a}$ and $\hat{a}$ are the bosonic creation and annihilation operators acting on the external state of each atom. The symbols $\eta_{\uparrow}=+1$ and $\eta_{\downarrow}=-1$ account for the opposite rotational directions for the two spin components.
%

Given by the Hamiltonian of the system, we construct the evolution operator in the following.
Since the single-particle Hamiltonians for different particles and different spins commute with each other, i.e., $[\hat H_l(t),\hat H_k(t)]=0~(l\neq k)$ and $[\hat H_{\uparrow}(t),\hat H_{\downarrow}(t)]=0$, the evolution operator for total evolution time $T$ can be formally written as:
\begin{equation}\label{Evo}
  \hat{U}(T)=\prod^N_{k=1} \hat{U}_k(T) ,\quad \hat{U}_k(T) =\hat{U}_\uparrow \ket{\uparrow} \bra{\uparrow}+ \hat{U}_\downarrow \ket{\downarrow} \bra{\downarrow},
\end{equation}
with $\hat{U}(T)$ and $\hat{U}_k(T)$ being the evolution operators of the system and individual particle, respectively.

If the multi-particle Hamiltonian is time-dependent, it does not commute with itself at different time, i.e., $[\hat{H}(t),\hat{H}(t')]\neq 0$.
Thus, one cannot evaluate the evolution operator by mere integration.
To solve this problem, we apply the Magnus expansion~\cite{Blanes2009}.
Suppose that the evolution operator of the system can be expressed as
\begin{equation}\label{EvoH}
  \hat{U}_{\sigma}(t)=\hat{U}_0(t) \hat{U}_{\sigma}^{I}(t),
\end{equation}
with
\begin{equation}\label{EvoH0}
  \hat{U}_0(t)=\exp \left(-i\frac{\hat{H}_0}{\hbar}t\right).
\end{equation}
Substitute Eqs.~\eqref{EvoH} and~\eqref{EvoH0} into the Schr\"{o}dinger equation, we can find out that the operator $\hat{U}_{\sigma}^{I}(t)$ satisfies
\begin{eqnarray}
\frac {\partial \hat{U}_{\sigma}^{I} (t) }{\partial t} &=&\left[ \hatd{U}_0(t) \frac{-i\hat{H}_{\sigma}^{I} }{\hbar} \hat{U}_0(t) \right] \hat{U}_{\sigma}^{I}(t)\nonumber \\
&=& A_{\sigma}(t)\left(e^{-i\omega t}\hat{a}-e^{i\omega t}\hatd{a}\right),
\end{eqnarray}
with
\begin{equation}\label{AA}
  A_{\sigma}(t)= \sqrt{\frac{m\omega}{2\hbar}} r \left[\omega_s+\eta_{\sigma}\omega_p(t)\right].
\end{equation}
Following the procedure of Magnus expansion, we can obtain
\begin{equation}\label{UI}
  \hat{U}_{\sigma}^{I}(T)=\exp\left(\alpha_{\sigma}^{*} \hat{a}-\alpha_{\sigma} \hatd{a}\right) \exp\left(i\phi_{\sigma}\right),
\end{equation}
where
\begin{equation}\label{alpha}
  \alpha_{\sigma} = \int^T_0 A_{\sigma}(t) e^{i\omega t} dt,
\end{equation}
and
\begin{equation}\label{phi}
  \phi_{\sigma} = \int^T_0 \int^{t_1}_0 A_{\sigma}(t_1)A_{\sigma}(t_2)\sin[\omega(t_1-t_2)] d{t_2} d{t_1}.
\end{equation}
Here, $T$ is the total evolution time and $T$ is determined by the choice of $\omega_p(t)$ satisfying $\int_{0}^{T}\omega_p(t)\text{d}t=\pi$.
Therefore, the final form of the evolution operator reads
\begin{equation}\label{UU}
  \hat{U}_{\sigma}=\exp\left(-i\omega t\hatd{a}\hat{a}\right)\exp\left(\alpha_{\sigma}^* \hat{a}-\alpha_{\sigma} \hatd{a}\right) \exp\left(i\phi_{\sigma}\right).
\end{equation}
The above expressions are general results for time-dependent parameter $\omega_p(t)$ and are valid for fixed parameter $\omega_p$ as well.

So far, we have derived the state-dependent evolution operator which characterizes the dynamics of the counter-rotation for the two spin components.
The evolution operator indicates that the effect of the rotation is equivalent to the displacement operator, the phase accumulation with eigen-frequency and the state-dependent phase shift.
By applying the evolution operator~\eqref{Evo} on the initial state, the output state after the free evolution becomes:
\begin{eqnarray}\label{output}
  \ket{\Psi(\omega_s)}_{out} &=& \hat{U} \ket{\Psi}_{in} \nonumber \\
  &=& \!\frac{1}{\sqrt{2}} \!\left[\bigotimes^N_{k=1} \!\left( \hat{U}_{\uparrow}\ket{\uparrow}\ket{0} \right)_k \!+\! \bigotimes^N_{k=1}\! \left(\hat{U}_{\downarrow}\ket{\downarrow}\ket{0} \right)_k \right].
\end{eqnarray}
Here, the information of the parameter $\omega_s$ is imprinted in the output state of the system $\ket{\Psi(\omega_s)}_{out}$.
Finally, we can acquire the value of $\omega_s$ from measuring the output state $\ket{\Psi(\omega_s)}_{out}$ and estimate the uncertainty of the parameter $\Delta\omega_s$.

\begin{figure}[!h]
\includegraphics[width=\columnwidth]{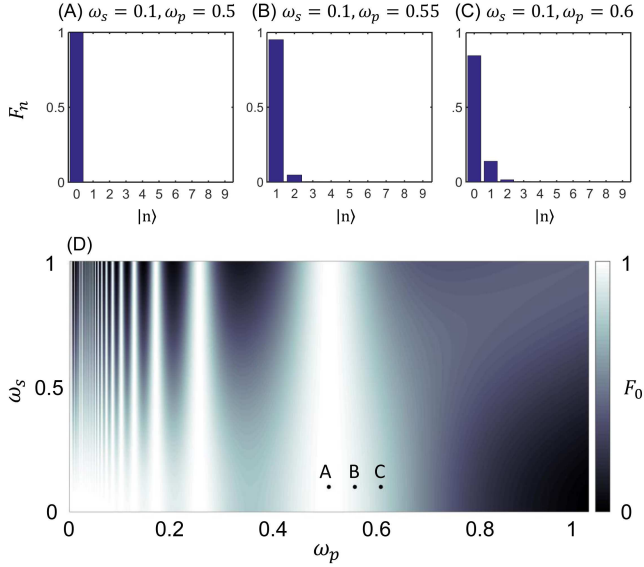}
\caption{\label{phase}(color online). The phase diagram of fidelity between its evolved external state and the initial ground state $F_0=|_k\langle\psi_{ex}|0\rangle_{k}|^2$. The distributions of fidelity $F_n=|_k\langle\psi_{ex}|n\rangle_{k}|^2$  for (A) $\omega_s=0.1,\omega_p=0.5$, (B) $\omega_s=0.1,\omega_p=0.55$ and (C) $\omega_s=0.1,\omega_p=0.6$ are shown. Here $\ket{0}$ is the external ground state of the potential well, and the $\ket{n}$ is the n-th excited state $(n\geq1)$. }
\end{figure}

At first, we vary both $\omega_s$ and $\omega_p$ in a wide range continuously, and calculate the corresponding output states.
Interestingly, we found that the external part of the output state is sensitively affected by the choice of $\omega_s$ and $\omega_p$.
For the $k$-th particle, different choice of $\omega_s$ and $\omega_p$ leads to different fidelity between the evolved external state $|\psi_{ex}\rangle_{k}$ and its initial ground state $F_0=|_k\langle\psi_{ex}|0\rangle_{k}|^2$.
The fidelity $F_0$ can characterize the probability of the atoms staying in the initial ground state of the external potential.
In Fig.~\ref{phase}~(d), we plot the phase diagram of the fidelity $F_0$ with $\omega_s, \omega_p \in[0,1]$.
For most $\omega_s$, the dark regions ($F_0\ll1$) alternate with the bright regions ($F_0\approx1$) as $\omega_p$ increase from 0 to 1. $F_0$ oscillates rapidly when $\omega_p$ is relatively small and a large bright area appear in the center of $\omega_p=0.5$.
In the brightest line where $F_0=1$ (e.g., $\omega_p=0.5$), the external state of every atom stays in the ground state all the time and the system's external state can be assumed to be unchanged during the Sagnac phase accumulation.
In general, it is beneficial to keep the external state unchanged during the interferometry and exploit the entanglement among the internal states of the atoms to perform high-precision measurement.

Obviously, the largest bright area is in the vicinity of $\omega_p=0.5$. We fix $\omega_s=0.1$ and choose $\omega_p=0.5, 0.55, 0.6$ to illustrate the distribution of fidelities projecting on $\ket{n}$, where $|0\rangle$ is the external ground state and $\ket{n}$ is the $n$-th external excited state $(n\geq1)$, see Fig.~\ref{phase}~(A)-(C).
For $\omega_p=0.5$, the external state keeps in $\ket{0}$. While for $\omega_p=0.55$ and $\omega_p=0.6$, $F_0$ decreases and the evolved external state has some components of other excited state $\ket{n}$.
It is shown that, one can specifically choose $\omega_p=0.5$ to measure the rotation frequency $\omega_s$ since the fidelity $F_0$ is very close to 1 in the vicinity of $\omega_p=0.5$. This can tolerate small deviation of $\omega_p$ around $0.5$ if some unavoidable errors exist.

Based on the phase diagram of Fig.~\ref{phase}~(D), one can carefully vary $\omega_p$ according to $\omega_s$ to ensure that $F_0\simeq1$.
Under this situation, the external state can be truncated only at the ground state.
In the following, we will show that, this situation is advantageous for the high-precision Sagnac interferometry and the Heisenberg-limited measurement can be achieved.
In addition, by using this truncation, some analytic results can be derived, see Sec.~\ref{Sec3} and~\ref{Sec4} .

\section{\label{Sec3}Ultimate Angular Frequency Measurement Precision}
In the framework of the quantum metrology, for a parameter-dependent output state, the uncertainty of the estimated parameter is limited by the QCRB:
\begin{equation}\label{QCRB}
  \Delta\omega_s\ge\Delta\omega_s^{Q}\equiv\frac{1}{\sqrt{\nu F_Q(\omega_s)}},
\end{equation}
where $\nu$ is the times of independent experiments and the uncertainty is defined as $\Delta\omega_s=\sqrt{\langle\omega_s^2\rangle-\langle\omega_s\rangle^2}$.
$F_Q(\omega_s)$ is the so-called quantum Fisher information (QFI), which can be expressed as a function of the output state $\ket{\Psi(\omega_s)}_{out}$ and its derivative with respect to the parameter $\omega_s$, i.e.,
\begin{equation}\label{QFI}
  F_Q(\omega_s)=4\left[ \braket{\Psi'(\omega_s)}{\Psi'(\omega_s)}\!-\!\begin{vmatrix} \braket{\Psi'(\omega_s)}{\Psi(\omega_s)}_{out} \end{vmatrix}^2\right],
\end{equation}
with $\ket{\Psi'(\omega_s)}=\text{d}\ket{\Psi(\omega_s)}_{out}/\text{d}\omega_s$.
The QFI determines the ultimate value of a parameter uncertainty for a given parameter-dependent output state.
The larger QFI $F_Q(\omega_s)$ corresponds to a smaller parameter uncertainty $\Delta\omega_s$.

In turns of our protocol, we first choose some specific values of $\omega_s$ and $\omega_p$ which satisfies $F_0 \approx 1$ according to Fig.~\ref{phase}~(D).
In this case, the external state can be truncated only at the ground state $\ket{0}$. Thereafter, the explicit expressions of the $\Delta\omega_s^Q$ and QFI $F_Q(\omega_s)$ could be evaluated.
The output state can be calculated as
\begin{eqnarray}
\ket{\Psi(\omega_s)}_{out}&\approx& \frac{1}{\sqrt{2}} \bigotimes^N_{k=1} \left[ \exp\left(-\frac{|\alpha_\uparrow|^2}{2} \right) \exp\left(i\phi_\uparrow\right)\ket{0}\ket{\uparrow}\right]_k \nonumber \\
&+& \frac{1}{\sqrt{2}} \bigotimes^N_{k=1} \left[\exp\left(-\frac{|\alpha_\downarrow|^2}{2} \right)\exp(i\phi_\downarrow)\ket{0} \ket{\downarrow} \right]_k \nonumber \\
&=& \frac{1}{\sqrt{2}}\!\left[ C_\uparrow^N \bigotimes^N_{k=1}\ket{\uparrow}_k\ket{0}_k \!+\! C_\downarrow^N \bigotimes^N_{k=1}\ket{\downarrow}_k\ket{0}_k \right].
\end{eqnarray}
Here, the coefficients $C_\uparrow$ and $C_\downarrow$ are respectively
\begin{eqnarray}
C_\uparrow & = &  \exp \left[-\frac {m r^2} {2 \hbar \omega}\left ( \omega_s + \omega_p \right)^2 \left ( \cos \frac {\pi \omega} {\omega_p} - 1 \right) \right]
    \nonumber \\
& &\times \exp \left[ i \frac {m r^2} {2\hbar \omega} \left (\omega_s + \omega_p \right)^2  \left ( {\pi} - \omega_p \sin \frac {\pi \omega_p} {\omega} \right) \right] ,
\end{eqnarray}

and
\begin{eqnarray}
C_\downarrow & = &  \exp \left[-\frac {m r^2} {2 \hbar \omega}\left ( \omega_s - \omega_p \right)^2 \left ( \cos \frac {\pi \omega} {\omega_p} - 1 \right) \right]
    \nonumber \\
& &\times \exp \left[ i \frac {m r^2} {2\hbar \omega} \left (\omega_s - \omega_p \right)^2  \left ( {\pi} - \omega_p \sin \frac {\pi \omega_p} {\omega} \right) \right] .
\end{eqnarray}
\begin{widetext}
Then, its derivative with respect to the $\omega_s$ is:
\begin{equation}
\ket{\Psi'(\omega_s)}= \frac{\text{d}\ket{\Psi(\omega_s)}_{out}}{\text{d} \omega_s} = \frac{1}{\sqrt{2}}\left[ \!N C_\uparrow^{N\!-\!1} C_\uparrow' \bigotimes^N_{k=1}\ket{\uparrow}_k\ket{0}_k \!+\! \!N C_\downarrow^{N\!-\!1} C_\downarrow' \bigotimes^N_{k=1}\ket{\downarrow}_k\ket{0}_k \right].
\end{equation}

Thus, we could calculate the QFI through Eq.~\eqref{QFI}, and its final expression can be written as

\begin{eqnarray} \label{QFI_ex}
F_Q (\omega_s) &=& \frac {m^2 N^2 r^4} {\hbar^2 \omega^2 \omega_p^2} \times
 \left[ \pi^2 \omega^2 + 2 \omega_p^2 - 2 \omega_p \left( \omega_p \cos \frac{\pi \omega}{\omega_p} + \pi \omega \sin \frac{\pi \omega}{\omega_p} \right) \right]
\nonumber \\
& & \times \left\{ 2 D_- ^N (\omega_p - \omega_s)^2 + 2 D_+^N (\omega_p + \omega_s)^2 - \left[  D_-^N (\omega_s - \omega_p) + D_+ ^N (\omega_s + \omega_p) \right]^2  \right\}.
\end{eqnarray}

\end{widetext}
In which, $D_+$ and $D_-$ are
\begin{eqnarray}
D_ + &=& e^{\frac {m r^2 (\omega_s + \omega_p)^2 (\cos \frac { \pi \omega} {\omega_p} - 1)} {\hbar \omega}} \nonumber \\
D_ - &=& e^{\frac {m r^2 (\omega_s - \omega_p)^2 (\cos \frac { \pi \omega} {\omega_p} - 1)} {\hbar \omega}}
\end{eqnarray}

Although Eq.~\eqref{QFI_ex} is lengthy, one can substitute the specific values of $\omega_s$ and $\omega_p$ (ensure that $F_0\approx1$, e.g., $\omega_s=0.1, \omega_p=0.5$) to simplify the expression.
Surprisingly, the QFI is exactly proportional to the square of total particle number, i.e.,
\begin{equation}\label{QFI-1}
  F_Q(\omega_s)\propto N^2.
\end{equation}

\begin{figure}[!h]
\includegraphics[width=\columnwidth]{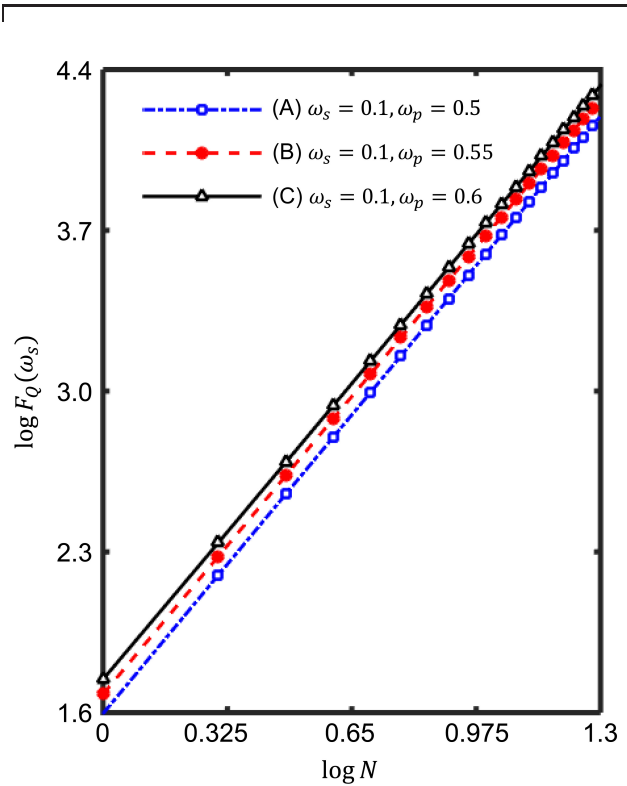}
\caption{\label{QFI_fig}(color online). The log-log scaling of QFI $F_Q(\omega_s)$ versus the total particle number $N$ under different choices of $\omega_s$ and $\omega_p$. The slopes of the lines are approximately equal to 2, which indicates $F_Q(\omega_s)\propto N^2$. To perform the numerical calculation, here we set $m=\omega=\hbar=1$.}
\end{figure}

Further, to verify the above analytic results, we follow the standard procedure and calculate the QFI numerically.
For $\omega_s=0.1$ and $\omega_p=0.5$, the numerical calculation agrees with the analytic result perfectly.
It confirms the validity of the truncation at the external ground state under this situation.
On the other hand, we choose other parameters that do not meet the condition of $F_0\approx1$.
For $\omega_s=0.1, \omega_p=0.55$ and $\omega_s=0.1, \omega_p=0.6$, the QFI also exhibit quadratic dependence on the total particle number.
The quadratic dependence is insensitive with the choices of $\omega_s$ and $\omega_p$.
In order to show the relation clearly, we give the log-log scaling of ${F_Q(\omega_s)}$ with respect to $N$ under the above three sets of $\omega_s$ and $\omega_p$, see Fig.~\ref{QFI_fig}.
The slopes of the three lines are all nearly 2, which confirms that $F_Q(\omega_s)\propto N^2$.
Therefore, the ultimate measurement precision can be derived,
\begin{equation}\label{HL}
    \Delta\omega_s^{Q} \propto \frac{1}{\sqrt{F_Q(\omega_s)}} \propto \frac{1}{N},
\end{equation}
which is inversely proportional to the total particle number, attaining the Heisenberg limit.
It is shown that, compared with the single-particle scheme~\cite{Fernholz2015}, the rotation measurement precision via Sagnac interferometry can be improved from SQL to Heisenberg limit by using the input GHZ state.

\section{\label{Sec4} Parity Measurement}

The ultimate measurement precision obtained via QFI is a theoretical bound which is independent on the choices of observable measurements.
In realistic scenarios, one would also be interested in how to approach the QCRB via certain achievable measurements.
For the maximally entangled state, parity measurement is assumed to be one of the effective candidates to saturate the QCRB and attain the Heisenberg limit~\cite{Gerry2003,Campos2003,Gerry2010}.

In our scheme, we also try to evaluate the rotation measurement precision via parity measurement.
The parity operator for $\ket{\downarrow}$ can be expressed as $\hat{P}=\exp\left[i\pi \left(\sum_{k=1}^N \ket{\downarrow}_k\bra{\downarrow}_k\right)\right]=\exp\left[i\pi (\frac{N}{2}-\sum_{k=1}^N \hat{\sigma}_z^{(k)})\right]$.
In Sec.\ref{Sec2}, we have presented the dynamical evolution and the output state~\eqref{output}.
Before the parity measurement, a $\pi/2$ pulse is implemented to rotate the output state for recombination, i.e, $\ket{\Psi}_f =\exp( -i \frac{\pi}{2} \hat{R} ) \ket{\Psi}_{out}$, where $\hat{R}=\frac{1}{2}\sum_{k=1}^N \hat{\sigma}_y^{(k)}$ with $\hat{\sigma}_y^{(k)}=\frac{1}{2i}\left(\ket{\downarrow}_k\bra{\uparrow}_k-\ket{\uparrow}_k\bra{\downarrow}_k\right)$.
The average of the parity measurement for $\ket{\downarrow}$ is written as $\left\langle{\hat{P}} \right\rangle =_f\bra{\Psi} \hat{P} \ket{\Psi}_f$ and the corresponding variance is given by $(\Delta \hat{P})^2=\langle \hat{P}^2 \rangle -\langle \hat{P} \rangle^2$.
Finally, from the results of parity measurement, the corresponding uncertainty of parameter $\omega_s$ can be estimated.
The standard deviation $\Delta\omega_s$ can be evaluated with the error propagation formula:
\begin{equation}\label{Deltaomega}
  \Delta\omega_s=\frac{\Delta \hat{P}}{|\partial \langle \hat{P} \rangle/\partial \omega_s|}.
\end{equation}


Still, we first apply the approximation that the external state is restricted in the ground state for specified range of $\omega_s$ and $\omega_p$ according to Fig.~\ref{phase}~(d).
Following the above procedure with the approximation, the final state after an additional $\pi/2$ pulse would be:
\begin{eqnarray}
\ket{\Psi}_f&=&\exp (-i\frac{\pi}{2}\hat{R})\hat{U}\ket{\Psi}_{in} \nonumber \\
&\approx& \frac{1}{\sqrt{2}} \bigotimes^N_{k=1} [(\frac{\ket{\uparrow}+\ket{\downarrow}}{2})\bigotimes\exp(-\frac{|\alpha_\uparrow|^2}{2})\exp(i\phi_\uparrow)\ket{0}]_k \nonumber \\
&+& \frac{1}{\sqrt{2}} \bigotimes^N_{k=1} [(\frac{\ket{\downarrow}-\ket{\uparrow}}{2})\bigotimes\exp(-\frac{|\alpha_\downarrow|^2}{2})\exp(i\phi_\downarrow)\ket{0}]_k. \nonumber\\
&&
\end{eqnarray}

\begin{widetext}
Applying the parity measurement $\hat P$ on the final state, the expectation value of the parity measurement can be obtained (hereafter, we set $\hbar=1$, $r=1$, $m=1$ and $\omega=1$),

\begin{eqnarray}\label{P1}
\langle \hat{P} \rangle &=& _f\bra{\Psi} \hat{P} \ket{\Psi}_f =  \frac { (-1)^N \cos \left[ 2 N \omega_s  \left(\omega_p \sin \frac {\pi } {\omega_p} - \pi  \right) \right]} {e^{\left[ N \left ( \omega_p^2 + \omega_s^2 \right) \left (1 - \cos \frac {\pi } {\omega_p} \right) \right]}},
\end{eqnarray}

\begin{eqnarray}\label{P2}
\langle \hat{P}^2 \rangle &=& _f\bra{\Psi} \hat{P}^2 \ket{\Psi}_f = \frac {e^ { N \left ( \omega_p - \omega_s \right)^2 \left (\cos \frac {\pi} {\omega_p} - 1 \right)} + e^{ N \left ( \omega_p + \omega_s \right)^2 \left (\cos \frac {\pi } {\omega_p} - 1 \right)}} {2}.
\end{eqnarray}
Meanwhile, the variance of the parity measurement for the final state can be also given by
\begin{eqnarray}\label{P3}
\left( \Delta \hat{P} \right)^2
&=&
\frac{
e^{N \left(\omega_s+\omega_p\right) ^2 \left( {\cos \frac{\pi}{\omega_p}}-1 \right ) }+e^{ N \left(\omega_s-\omega_p\right) ^2 \left( {\cos \frac{\pi}{\omega_p}}-1 \right )}
}{2} -   \frac{ \cos^2  \left[ 2N \omega_s \left(  \omega_p \sin \frac{\pi } { \omega_p}-\pi  \right)  \right] }
{e^{2 N \left( \omega_s^2 +\omega_p^2 \right) \left(   1-\cos \frac{\pi}{\omega_p}  \right) }}.
\end{eqnarray}
Eventually, we can obtain the standard deviation of the $\omega_s$
\begin{eqnarray}\label{P4}
\Delta \omega_s&=& \frac{\Delta \hat{P}}{|\partial \langle \hat{P} \rangle /\partial \omega_s|} \nonumber \\
&=& (-1)^N e^{N \left( \omega_p^2 +\omega_s^2 \right) \left(1- \cos \frac{\pi}{\omega_p}\right)} \nonumber \\
&  & \times \frac{ \sqrt{e^{N\left ( \omega_p- \omega_s \right)^2\left ( \cos \frac {\pi} {\omega_p} -1 \right)}+e^{N\left ( \omega_p+ \omega_s \right)^2 \left ( \cos\frac {\pi} {\omega_p} -1 \right)}-2 e^{2 N\left ( \omega_p^2+ \omega_s^2 \right) \left ( \cos\frac {\pi}{\omega_p} -1\right) } \cos^2 {\left[2 N\omega_s\left ( \pi- \omega_p\sin {\frac {\pi} {\omega_p}} \right) \right]}}}
{2\sqrt {2} \left|  N  \omega_s\left (\cos\frac {\pi} { \omega_p} - 1 \right) \cos \left[ 2 N\omega_s\left ( \pi - \omega_p \sin\frac {\pi} {\omega_p} \right) \right] + \left ( \omega_p \sin\frac {\pi} {\omega_p} - \pi \right) \sin \left[ 2 N \omega_s\left (\pi - \omega_p \sin \frac {\pi} {\omega_p} \right) \right] \right |}.
\end{eqnarray}

\end{widetext}

Here, one can easily observe that, for $\omega_p \approx\frac{1}{L}$ [$L$ is an integer larger than 1, this corresponds to some of the brightest vertical lines in Fig.~\ref{phase}~(d)], $\cos \frac{\pi}{\omega_p}-1\approx 0$Œ, and $\sin\frac {\pi}{\omega_p}\approx 0 $, Eqs.~\eqref{P1}-\eqref{P4} could be further simplified to

\begin{eqnarray}\label{P5}
\langle \hat{P} \rangle
&=&  (-1)^{N} \cos \left( 2N \pi \omega_s  \right),
\end{eqnarray}

\begin{eqnarray}\label{P6}
\langle \hat{P}^2 \rangle = 1,
\end{eqnarray}

\begin{eqnarray}\label{P7}
\left( \Delta \hat{P} \right)^2
&=& \sin \left( 2N \pi \omega_s  \right),
\end{eqnarray}
and
\begin{eqnarray}\label{P8}
\Delta \omega_s= \frac{1}{2\pi N}.
\end{eqnarray}

From Eq.~\eqref{P8}, it is clearly shown that the deviation of the $\omega_s$ is exactly proportional to the particle number, i.e.,$\Delta \omega_s \propto N^{-1}$.
This is consistent with the ultimate bound predicted with QFI.
In addition, the deviation $\Delta\omega_s$ is independent of $\omega_s$ itself so that the deviation is determined by the choice of $\omega_p$.
The best measurement precision can be achieved when $\omega_p=0.5$.

\begin{figure}[!htb]
\includegraphics[width=\columnwidth]{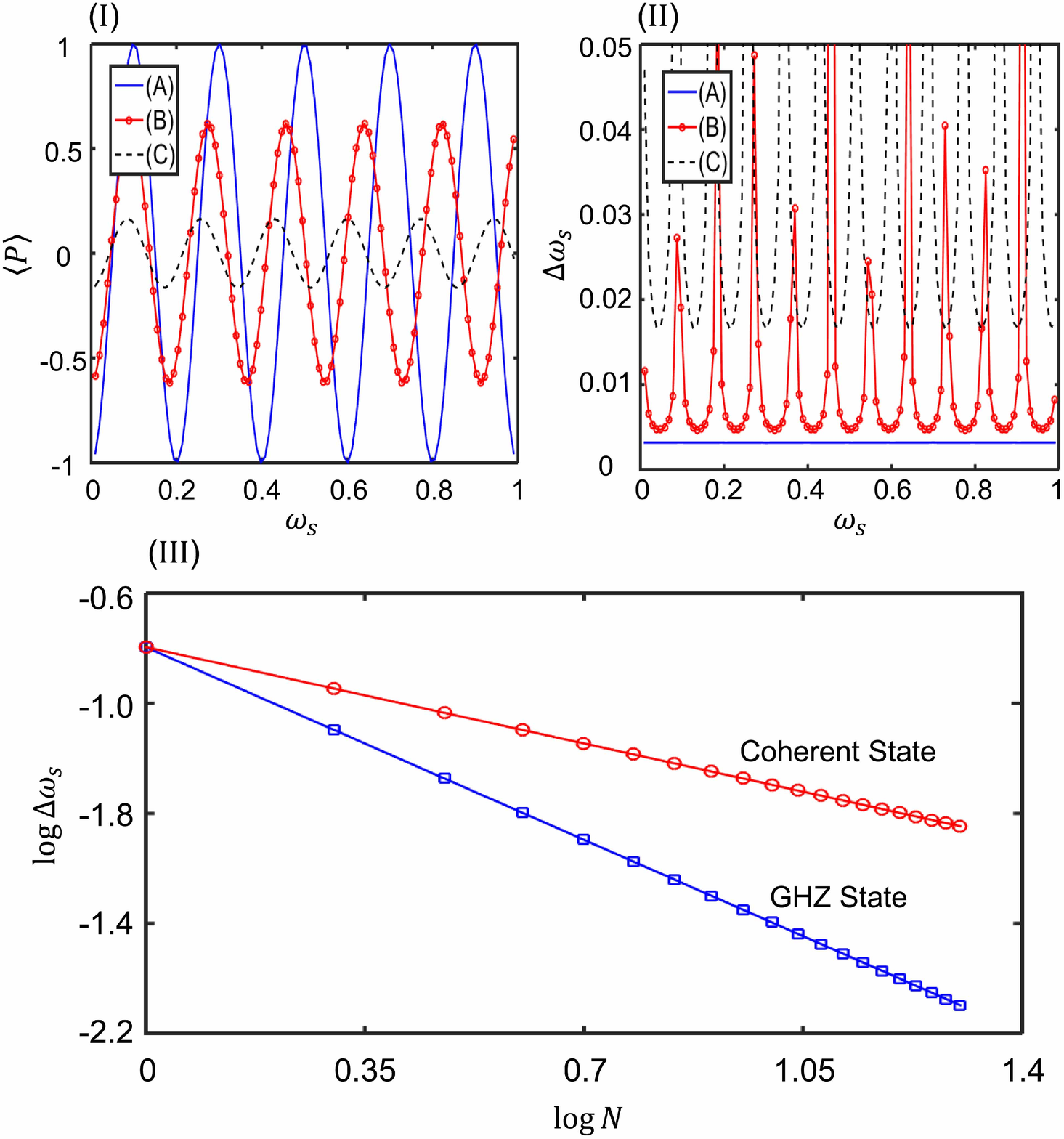}
\caption{\label{ParRes}(color online). Numerical results of the parity measurement. (I) The expectation value of the parity is a sinusoidal function with respect to $\omega_s$. (II) The dependence of the measurement precision on $\omega_s$ itself. For (I) and (II), (A)$\omega_s=0.1,\omega_p=0.5$, (B)$\omega_s=0.1,\omega_p=0.55$, and (C) $\omega_s=0.1,\omega_p=0.6$, and the total particle number $N=5$. (III) For case (A), the log-log scaling of QFI versus $N$ with input GHZ state and spin coherent state are shown, respectively. }
\end{figure}

Meanwhile, we check the above analytic results by numerical calculation.
The numerical results via parity measurement for $N=5$ are presented in the Fig.\ref{ParRes}~(I) and~(II).
For $\omega_p=0.5$, The expectation values of parity oscillate sinusoidally from $-1$ to $1$ with respect to $\omega_s$.
The deviation $\Delta\omega_s$ is a horizontal line versus $\omega_s$, which is independent of $\omega_s$.
This result perfectly agrees with the analytic Eq.~\eqref{P8}.
On the other hand, we also evaluate $\Delta\omega_s$ via parity measurement with $\omega_p=0.55$ and $\omega_p=0.6$.
The contrast of $\langle \hat{P}\rangle$ drops rapidly as $\omega_p$ becomes far away $0.5$.
The period of the sinusoidal oscillation also changes.
These result in the reduction of the measurement precision.
Based on our numerical calculation, we confirm that the best standard deviation of $\omega_s$ can be achieved under the choice of $\omega_p=0.5$.

Finally, we fix $\omega_p=0.5$ and evaluate the log-log scaling of $\Delta\omega_s$ versus the total particle number $N$, see Fig.~\ref{ParRes}~(III).
Compared with the input coherent spin state (which is equivalent to the single-particle scheme), the dependence of the deviation $\Delta\omega_s$ on $N$ based upon our scheme is quadratic rather than linear.
%
Thus, by implementing the input GHZ state for Sagnac interferometry, the parity measurement is an optimal way to saturate the ultimate precision bound and can be used to perform high-precision rotation sensing at the Heisenberg limit.

\section{\label{Sec5}Discussion and Conclusion}
We have shown how many-body entanglement enhances the rotation sensing.
The maximally entangled state (GHZ state) can be prepared by various methods~\cite{You2003, Zhang2003, Lee2006}.
For Bose condensed atoms, the GHZ state can be created by dynamical nonlinear evolution~\cite{Pawlowski2013,Gross2012} or adiabatic ground state preparation~\cite{Lee2006,Gross2012}.
By driving internal state Raman transitions via laser pulses~\cite{You2003} or classical fields~\cite{Zhang2003}, an $N$-GHZ state can also be effectively generated in spin-1 Bose-Einstein condensates.
In addition, the spin-dependent control is another important element in our scheme, which may be realized by adiabatic dressed potentials.
State-dependent control of atomic transport in the toroidal trap has been proposed~\cite{Fernholz2007}.
The RF dressed potential and the coherent control atomic motion have been demonstrated in experiments~\cite{Treutlein2004,Schumm2005,Morizot2006,Sherlock2011,Szmuk2015}.
These techniques could be applied to perform the state-dependent control of atomic transport on the ring-shaped traps~\cite{Fernholz2016}.
On the other hand, our Sagnac interferometry scheme could be applied to other many-body systems, such as trapped ions~\cite{Leibfried2003}.
A protocol for using trapped ions to measure rotations via repeated round-trips Sagnac interferometry is proposed recently~\cite{Wes2016}.
By using the GHZ state~\cite{Blatt2011}, the rotation sensitivity can also be improved.

In summary, we have presented a multi-atom Sagnac interferometer scheme with maximally entangled state, which can attain the Heisenberg limit.
During the Sagnac phase accumulation, the internal and external states of the system are coupled with each other.
The uncertainty of the estimated angular frequency $\omega_s$ is sensitively influenced by the choice of the induced angular frequency $\omega_p$.
By optimally selecting $\omega_p$, the external state would stay in its initial ground state during the phase accumulation and the ultimate angular frequency measurement precision can reach the Heisenberg limit.
Further, we analyzed the angular frequency estimation via the parity measurement.
We found that the parity measurement may attain the Heisenberg limit imposed by the quantum Fisher information.
Our scheme can be extended to other kinds of many-body entangled states such as spin squeezed states~\cite{Kitagawa1993}, spin cat states~\cite{Huang2015} or twin Fock state~\cite{Campos2003}.
This may open up a way to perform high-precision rotation measurement with many-body quantum entanglement beyond the standard single-particle Sagnac effect.

\section*{\label{Sec6}Acknowledgement}
The author Chengyi Luo would like to thank Sheng Zhang, Zui Tao and Jingjia Chen for helpful discussions.
This work is supported by the National Basic Research Program of China (Grant No. 2012CB821305) and the National Natural Science Foundation of China (Grant No. 11374375, 11574405).

\end{document}